\documentclass[journal]{IEEEtran}
\usepackage{mathrsfs}
\usepackage{multirow}
\usepackage{booktabs}
\usepackage{hyperref}
\usepackage{color}
\usepackage{algorithm} %format of the algorithm
\usepackage{algorithmic} %format of the algorithm

\usepackage{flushend}

\usepackage{mathrsfs}

\usepackage{colortbl}
\definecolor{indian1}{rgb}{0,0,.6125}
\definecolor{indian2}{rgb}{0,0.0769,1}
\definecolor{indian3}{rgb}{0,0.6154,1}
\definecolor{indian4}{rgb}{0,0.0769,1}
\definecolor{indian5}{rgb}{0.5385,1,0.9231}
\definecolor{indian6}{rgb}{1,0.9230,0}
\definecolor{indian7}{rgb}{1,0.4615,0}
\definecolor{indian8}{rgb}{0.923,0,0}
\definecolor{indian9}{rgb}{0,0.2308,0}
\definecolor{indian10}{rgb}{0,	0.692307692307692,	0}
\definecolor{indian11}{rgb}{0.153846153846154,	0,	0.846153846153846}
\definecolor{indian12}{rgb}{0.615384615384615,	0,	0.384615384615385}
\definecolor{indian13}{rgb}{0,	0,	1}
\definecolor{indian14}{rgb}{0.538461538461538,	0,	1}
\definecolor{indian15}{rgb}{1,	0,	1}
\definecolor{indian16}{rgb}{1,	0,	0.538461538461538}

\definecolor{paviaU1}{rgb}{0,	0,	0.571428571428571}
\definecolor{paviaU2}{rgb}{0,	0.428571428571429,	1}
\definecolor{paviaU3}{rgb}{0.428571428571429,	1,	0.571428571428571}
\definecolor{paviaU4}{rgb}{1,	0.714285714285714,	0}
\definecolor{paviaU5}{rgb}{0.857142857142857,	0,	0}
\definecolor{paviaU6}{rgb}{0,	0.714285714285714,	0}
\definecolor{paviaU7}{rgb}{0.428571428571429,	0,	0.571428571428571}
\definecolor{paviaU8}{rgb}{0.142857142857143,	0,	1}
\definecolor{paviaU9}{rgb}{1,	0,	1}

\ifCLASSINFOpdf
  \usepackage[pdftex]{graphicx}
  \graphicspath{{../pdf/}{../jpeg/}}
\else
  \usepackage[dvips]{graphicx}
  \graphicspath{{../eps/}}
\fi

\usepackage{algorithm} %format of the algorithm
\usepackage{algorithmic} %format of the algorithm
\usepackage{amsmath,amsthm,amssymb}

\begin{document}
%\title{S3RN: spatial-spectral Super-Resolution Network for Hyperspectral Imagery}
\title{Learning Spatial-Spectral Prior for Super-Resolution of Hyperspectral Imagery}

\author{Junjun~Jiang,~\IEEEmembership{Member,~IEEE,~}
        He Sun,
        Xianming Liu,~\IEEEmembership{Member,~IEEE,~}
        and Jiayi Ma,~\IEEEmembership{Member,~IEEE}% <-this % stops a space

\IEEEcompsocitemizethanks{
%\IEEEcompsocthanksitem M. Shell is with the Department
%of Electrical and Computer Engineering, Georgia Institute of Technology, Atlanta,
%GA, 30332.\protect\\
%E-mail: see http://www.michaelshell.org/contact.html
\IEEEcompsocthanksitem Junjun Jiang and Xianming Liu are with the School of Computer Science and Technology, Harbin Institute of Technology, Harbin 150001, China, and are also with the Peng Cheng Laboray, Shenzhen, China. E-mail: \{jiangjunjun, csxm\}@hit.edu.cn.
\IEEEcompsocthanksitem He Sun is with the School of Computer Science and Technology, Harbin Institute of Technology, Harbin 150001, China. E-mail: 19s103179@hit.edu.cn.
\IEEEcompsocthanksitem Jiayi Ma is with the Electronic Information School, Wuhan University, Wuhan 430072, China. E-mail: jyma2010@gmail.com.
%\IEEEcompsocthanksitem Copyright (c) 2013 IEEE. Personal use of this material is permitted. However, permission to use this material for any other purposes must be obtained from the IEEE by sending a request to pubs-permissions@ieee.org. \protect\\
}% <-this % stops an unwanted space
\thanks{The research was supported by the National Natural Science Foundation of China (61971165, 61922027, 61773295), and also is supported by the Fundamental Research Funds for the Central Universities.} %
%\thanks{Copyright (c) 2016 IEEE. Personal use of this material is permitted. However, permission to use this material for any other purposes must be obtained from the IEEE by sending an email to pubs-permissions@ieee.org.}
}

%\markboth{}%
\markboth{IEEE Transactions on Computational Imaging,~Vol.~XXX, No.~XXX, XXX~2020}%
{Shell \MakeLowercase{\textit{\emph{et al.}}}: Bare Demo of IEEEtran.cls for Journals}



\maketitle

\begin{abstract}
Recently, single gray/RGB image super-resolution reconstruction task has been extensively studied and made significant progress by leveraging the advanced machine learning techniques based on deep convolutional neural networks (DCNNs). However, there has been limited technical development focusing on single hyperspectral image super-resolution due to the high-dimensional and complex spectral patterns in hyperspectral image. In this paper, we make a step forward by investigating how to adapt state-of-the-art residual learning based single gray/RGB image super-resolution approaches for computationally efficient single hyperspectral image super-resolution, referred as SSPSR. Specifically, we introduce a spatial-spectral prior network (SSPN) to fully exploit the spatial information and the correlation between the spectra of the hyperspectral data. Considering that the hyperspectral training samples are scarce and the spectral dimension of hyperspectral image data is very high, it is nontrivial to train a stable and effective deep network. Therefore, a group convolution (with shared network parameters) and progressive upsampling framework is proposed. This will not only alleviate the difficulty in feature extraction due to high dimension of the hyperspectral data, but also make the training process more stable. To exploit the spatial and spectral prior, we design a spatial-spectral block (SSB), which consists of a spatial residual module and a spectral attention residual module. Experimental results on some hyperspectral images demonstrate that the proposed SSPSR method enhances the details of the recovered high-resolution hyperspectral images, and outperforms state-of-the-arts. \textcolor[rgb]{1.00,0.00,0.00}{The source code is available at \url{https://github.com/junjun-jiang/SSPSR}}.
\end{abstract}


\section{Introduction}
\label{sec:intro}
Unlike human eyes, which can only be exposed to visible light, hyperspectral imaging  is an imaging technique for collection and processing information across the entire range of electromagnetic spectrum \cite{rickard1993hydice}. The most important feature of hyperspectral imaging is the combination of imaging technology and spectral detection technology. While imaging the spatial features of the target, each spatial pixel in a hyperspectral image is dispersed to form dozens or even hundreds of narrow spectral bands for continuous spectral coverage. Therefore, hyperspectral images have a strong spectral diagnostic capability to distinguish materials that look similar for humans.

However, the hyperspectral imaging system is often compromised due to the limitations of the amount of the incident energy. There is always a tradeoff between the spatial and spectral resolution of the real imaging process. With the increase of spectral features, if all other factors are kept constant to ensure a high signal-to-noise ratio (SNR), the spatial resolution will inevitably become a victim. Therefore, how to obtain a reliable hyperspectral image with high-resolution still remains a very challenging problem.

Super-resolution reconstruction can infer a high-resolution image from one or sequential observed low-resolution images \cite{park2003super}. It is a post-processing technique that does not require hardware modifications, and thus could break through the limitations of the imaging system. According to whether the auxiliary information (such as panchromatic, RGB, or multispectral image) is utilized, hyperspectral image super-resolution techniques can be divided into two categories: \emph{fusion based hyperspectral image super-resolution} (sometimes called hyperspectral image pansharpening) and \emph{single hyperspectral image super-resolution} \cite{yokoya2017hyperspectral}. The former merges the observed low-resolution hyperspectral image with the higher spatial resolution auxiliary image to improve the spatial resolution of the observed hyperspectral image. These fusion approaches based on Bayesian inference, matrix factorization, sparse representation, or recently advanced deep learning techniques have flourished in recent years and achieved considerable performance \cite{wei2015bayesian, yokoya2011coupled, akhtar2014sparse}. However, most of these methods all assume that the input low-resolution hyperspectral image and the high-resolution auxiliary image are well co-registered. In practical applications, obtaining such well co-registered auxiliary images would be difficult, if not impossible \cite{chen2015sirf, pan2018multispectral, Zhou2019AN}.

Compared with fusion based hyperspectral image super-resolution, single hyperspectral image super-resolution has received less attention and there has been limited advancement due to the spectral patterns in hyperspectral images and no additional auxiliary information. To exploit the abundant spectral correlations among successive spectral bands, several single hyperspectral image super-resolution approaches based on sparse and dictionary learning or low-rank approximation have been developed \cite{huang2014super, he2016super, wang2017hyperspectral, irmak2018map}. However, these hand-crafted priors can only reflect the characteristics of one aspect of the hyperspectral data.

Recently, deep convolutional neural network (DCNN) has shown extraordinary capability of modelling the relationship between the low-resolution images and high-resolution ones, \emph{i.e.}, single gray/RGB image super-resolution task \cite{dong2015image, lim2017enhanced, zhang2018image}. The practiced rationale in these schemes can be summarized as follows: given a very large number of example pairs of original images and their corrupted versions, a deep network can be learned to restore the degraded image to its source.

Specifically, compared with the single gray/RGB image super-resolution based on deep learning, in the single hyperspectral image super-resolution task, it is nontrivial to train a computationally efficient and effective deep network. This is mainly due to the following reasons: on the one hand, hyperspectral images are not as popular as natural images, the training sample number of available hyperspectral image dataset is extremely small. Even if we can collect a lot of images, hyperspectral images may be obtained by different hyperspectral cameras. The differences in the number of spectral bands and imaging conditions will make it more difficult to establish a unified deep network. On the other hand, the spectral dimensionality of hyperspectral image data itself is very high. Unlike traditional gray/RGB images, hyperspectral images often have hundreds of contiguous spectral bands, which calls for larger dataset to guarantee the training process. Otherwise, it is easy to cause the over-fitting problem.

In order to deal with the above problems caused by the lack of data and the inability to fully exploit the spatial information and spatial correlation characteristics in hyperspectral data, a \emph{group convolution} (with shared network parameters) and \emph{progressive upsampling} framework is proposed in this paper, which can greatly reduce the size of the model and make it feasible to obtain stable training results under small data conditions.
For exploiting the spatial and spectral correlation characteristics of hyperspectral data, we carefully design the spatial-spectral prior network (SSPN), which cascades multiple spatial-spectral blocks (SSBs). For each SSB, it contains a spatial residual module and a spectral attention residual module. The former consists of a standard residual block which is used to exploit spatial information of the hyperspectral data, while the latter consists of a spectral attention residual module which is used to extract spectral correlations. Through short and long skip connections, a residual in residual architecture is formed, which makes the spatial-spectral feature extraction more efficient.

\begin{figure*}[t]
  \centering
  \includegraphics[width=18cm]{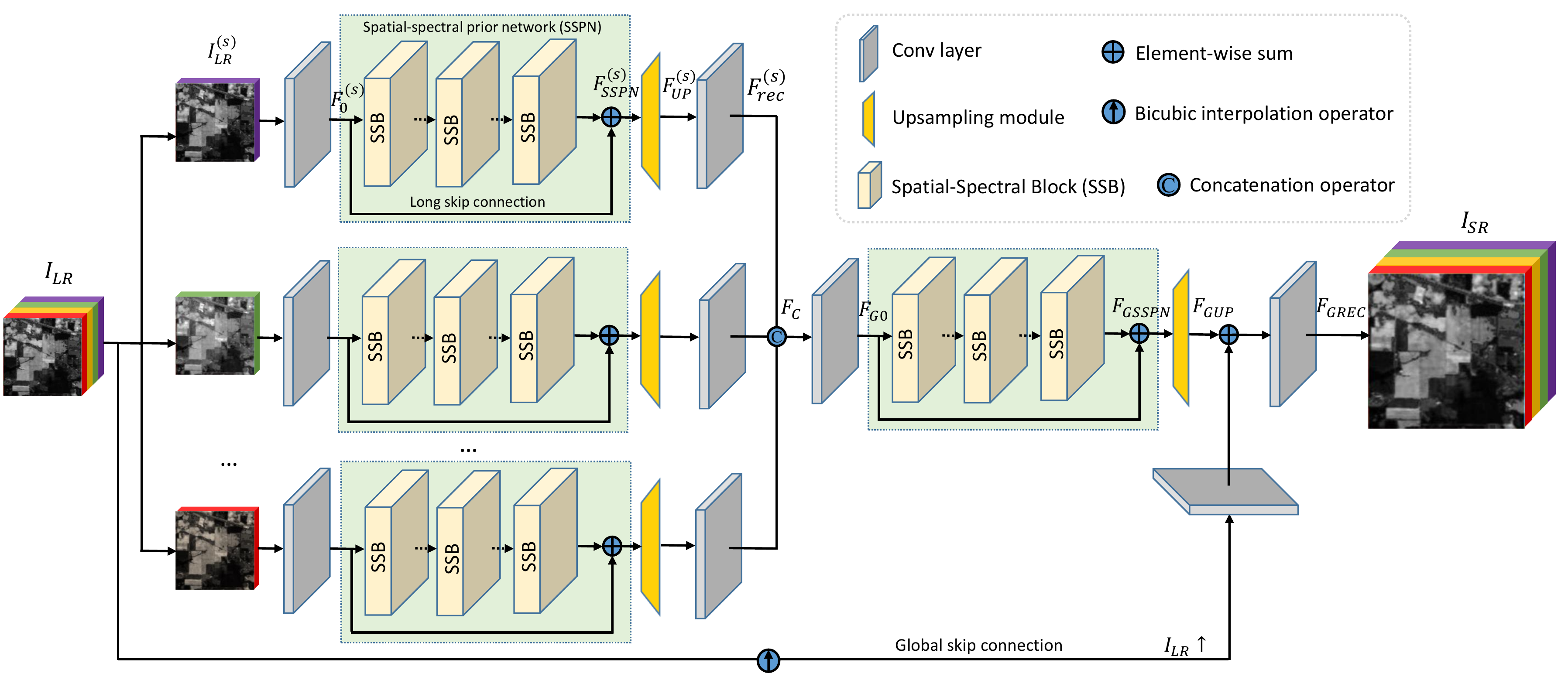}\\
  \caption{The overall network architecture of the proposed SSPSR network.}\label{fig:N5new}
\end{figure*}

Figure \ref{fig:N5new} shows the network architecture of our spatial-spectral prior network based super-resolution network (SSPSR). The input low-resolution hyperspectral image is firstly divided into several overlap groups. For each group, a branch network is applied to extract the spatial-spectral features of the input grouped hyperspectral images (a subset of the entire hyperspectral linages) and upscale them with a smaller unsampling factor (compared with the final target). And then, the output features of all branches are concatenated and fed to the following global spatial-spectral feature extraction and upsampling networks. Note that in order to let the SSPN in branch network and global network share the same structure, we insert a ``reconstruction'' layer after each branch upsampling module. Similar to many previous super-resolution networks, we also adapt a global residual structure to facilitate the prediction of the target. Therefore, in the proposed SSPSR network, the transmission of information flow is very flexible by designing these short (refer to residual spatial/spectral blocks), long (refer to the spatial-spectral prior network), global skip links.
During the training phase, we share the network parameters of each branch across all groups, which avoids heavy computational cost and simplifies the complex optimization process. Comprehensive ablation studies demonstrate the effectiveness of each component and the fusion strategy used in the proposed method. Comparison results with state-of-the-art single hyperspectral image super-resolution methods on two public datasets demonstrate the effectiveness of the proposed SSPSR network.

We summarize the main contributions of this paper as follows. Considering the limited hyperspectral training samples and the high dimensionality of spectral bands, it is difficult to learn the mapping relationship from low-resolution space to high- resolution space in one-step upsampling. Inspired by the idea of some general image super-resolution methods, which con- duct super-resolution progressively, we apply the progressive upsampling scheme to the single hyperspectral image super-resolution task and verify its effectiveness. In addition, we propose a spectral grouping and parameter sharing strategy to greatly reduce the parameters of the model and alleviate the difficulty in feature extraction. Inspired by the efficient residual learning and attention mechanism, we develop a spatial-spectral feature extraction network to fully exploit the spatial-spectral prior of hyperspectral images.

The rest of this paper is organized as follows: Section \ref{sec:related} presents the related work of hyperspectral image super-resolution. In Section \ref{sec:Proposed}, we give the details of our SSPSR network architecture and the SSB. Then, the network configuration and experimental results including ablation analysis are reported in Section \ref{sec:Experiments}. Finally, some conclusions are drawn in Section \ref{sec:Conclusions}.

\section{Related Work}
\label{sec:related}
In this section, we briefly review some methods that are most relevant to our work, which include \emph{fusion based hyperspectral image super-resolution}, \emph{single hyperspectral image super-resolution}, and \emph{single gray/RGB image super-resolution}. \emph{A list of hyperspectral image super-resolution resources collected by Jiang can be found at \cite{Jiang2020GithubHSI}}.

\subsection{Fusion based Hyperspectral Image Super-Resolution}
Remote sensing image fusion is a very challenging problem with long history. Generally speaking, this problem can be classified to two categories, pansharpening and super-resolution. In order to improve the spatial resolution of the multispectral images, some previous works cast the fusion problem into a variational reconstruction task by blending a panchromatic image with higher resolution. This is often referred as pansharpening. A taxonomy of pansharpening based fusion methods can be found in the literature \cite{aiazzi2012twenty,  he2014new, Li2019DDLPS, ma2020pan}.

Recently, low-resolution hyperspectral image and high-resolution multispectral image fusion based spatial resolution improvement technique, which is often referred as hyperspectral image super-resolution, has received extensive attention. For example, Yokoya \emph{et al.} \cite{yokoya2011coupled} proposed a coupled nonnegative matrix factorization (CNMF) based approach to infer the high-resolution hyperspectral images with a pair of high-resolution multispectral image and low-resolution hyperspectral image. To exploit the redundancy and correlation in spectral domain, some approaches have been proposed by exploiting the sparsity \cite{akhtar2015bayesian}, non-local similarity \cite{dong2016hyperspectral, xu2019nonlocal}, superpixel-guided self-similarity \cite{han2018self}, clustering manifold structure \cite{zhang2018exploiting}, tensor and low-rank constraints \cite{veganzones2015hyperspectral, dian2019hyperspectral}. Most recently, some deep learning based methods have gradually become popular due to its superior performance and fewer assumptions regarding the image prior \cite{yang2017pannet, dian2018deep, qu2018unsupervised, Borsoi2020TIP}. Inspired by the iterative optimization based on the observation model, some deep unfolding network for fusion based hyperspectral image super-resolution methods are becoming popular in recent years \cite{xie2019multispectral, Bihan2018DEEPCASD, Deng2020TIP}. The common idea of the above fusion based hyperspectral image super-resolution methods is to borrow high-frequency spatial information from high-resolution auxiliary image, and fuse these information to the target high-resolution hyperspectral image. Though these approaches have achieved very good performance, the major drawback of them is that a well co-registered auxiliary image with a higher resolution is needed. However, obtaining such a well co-registered auxiliary image would be arduous, if not impossible in practical applications \cite{chen2015sirf, pan2018multispectral, Zhou2019AN}.

\subsection{Single Hyperspectral Image Super-Resolution}
Without co-registered auxiliary image, single hyperspectral image super-resolution methods have still attracted considerable attention in reality. The pioneer work is proposed by Akgun \emph{et al.} \cite{akgun2005super}, in which a hyperspectral image acquisition model and the projection onto convex sets (POCS) algorithm \cite{bauschke1996projection} is applied to reconstruct the high-resolution hyperspectral image. By incorporating the low-rank and group-sparse constraints, Huang \emph{et al.} \cite{huang2014super} developed a novel method to tack with the unknown blurring problem. Recently, variants of sparse representations and dictionary learning based approaches are widely studied \cite{wang2017hyperspectral, li2016hyperspectral}. However, these methods have some drawbacks. First, they usually need to solve some complex and time consuming optimization problems in the testing phase. Second, the image priors are often hand-crafted and based on the internal example without consideration of any external information from external samples. Due to the superior performance in many computer vision problems, deep learning techniques have also been introduced into the single hyperspectral image super-resolution task very recently. For example, Yuan \emph{et al.} \cite{yuan2017hyperspectral} and Xie \emph{et al.} \cite{xie2019hyperspectral} firstly super-resolved the hyperspectral image based on the DCNNs, and then applied the nonnegative matrix factorization (NMF) to guarantee the spectral characteristic for the intermediate results. Essentially, they utilized DCNNs and matrix factorization to exploit the spatial and spectral features, separately, in a non-end-to-end manner. In \cite{mei2017hyperspectral}, Mei \emph{et al.} introduced a 3D full convolutional neural network to extract the feature of hyperspectral images. Although 3D convolution can well exploit the spectral correlation, the computational complexity is very large. Li \emph{et al.} \cite{li2018single} proposed a grouped deep recursive residual network (GDRRN) by designing a group recursive module and embedding it into a global residual structure. This group-wise convolution and recursive structure can guarantee that it could yield very good performance. In our previous work \cite{sun2019hyperspectral}, a feature pyramid block is designed to extract multi-scale features of the hyperspectral images. Most recently, inspired by the work of \cite{ulyanov2018deep}, which states that the image prior can be found within a CNN itself, Sidorov \emph{et al}. \cite{sidorov2019deep} developed an effective single hyperspectral-image restoration algorithm. In general, these deep methods achieve better results than traditional methods. However, due to the limited hyperspectral training samples and the high dimensionality of spectral bands, it is difficult to fully exploit the spatial information and the correlation among the spectra of the hyperspectral data.

\subsection{Single Gray/RGB Image Super-Resolution}
Recently, DCNN based approaches have achieved excellent performance over the single gray/RGB image super-resolution problem. The seminal work by Dong \emph{et al.} \cite{dong2015image} proposes a three layer convolutional neural network for the end-to-end image super-resolution(SRCNN) and achieved much better performance over conventional non-deep learning based methods. Benefiting from the residual learning, in VDSR \cite{kim2016accurate} and DRCN \cite{kim2016deeply} Kim \emph{et al.} introduced very deep network for image super-resolution and achieved better results than the three layer SRCNN. The residual structure was then adopted in LapSRN \cite{lai2017deep}, DRRN \cite{tai2017image}, and EDSR \cite{lim2017enhanced}. By simply attaching residual blocks, introducing the feedback, or incorporating non-local operations into a recurrent neural network, RDN \cite{zhang2018residual}, DBPN \cite{haris2018deep}, and NLRN \cite{LiuD2018Nips} are proposed. Inspired by the SE block \cite{Hu2017Squeeze}, Zhang \emph{et al}. developed a very deep network named RCAN by incorporating the channel attention module \cite{zhang2018image}. Most recently, Dai \emph{et al}. introduced the non-local block and presented a second-order attention network (SAN) to capture the long-range dependencies \cite{dai2019second}. Although fascinating results have been achieved, these methods are designed for the gray/RGB images, which have only one or three channels. When directly applying these approaches to the hyperspectral image, they will neglect the spectral correlations among spectra of the hyperspectral data, hindering the representation capacity of the network. In addition, for single gray/RGB image super-resolution, when using one- or three-channel pictures as network input, in order to extract features, a feature map of 64 (or more) channels is usually used. Similarly, if we also apply this 20-fold (or more) parameter growth network design scheme to hyperspectral images which have hundreds of channels, it will lead to a sharp increase in parameters. However, there is not enough hyperspectral data to support the model training like for the gray/RGB images.

\section{The Proposed SSPSR Method}
\label{sec:Proposed}
\subsection{Network Architecture}
In Fig. \ref{fig:N5new}, we show the network architecture of the proposed SSPSR method. It mainly consists of two parts: the branch networks and global network. For each branch network or the global network, it includes shallow feature extraction, spatial-spectral deep feature extraction, upsampling module, and reconstruction part. We denote $I_{LR} \in \mathbb{R}^{h\times w \times C}$ the input low-resolution hyperspectral image, $I_{SR} \in \mathbb{R}^{H\times W\times C}$ the corresponding output high-resolution hyperspectral image, and $I_{HR} \in \mathbb{R}^{H\times W \times C}$ the ground truth (original high-resolution hyperspectral image) of the input image $I_{LR}$. Our goal is to predict the high-resolution hyperspectral image $I_{SR}$ from the input low-resolution hyperspectral image $I_{LR}$ by the proposed end-to-end super-resolution reconstruction network,
\begin{equation}\label{eq:one}
I_{SR} = H_{Net}(I_{LR}),
\end{equation}
where $H_{Net}(\cdot)$ denotes the function of the proposed SSPSR method.

Different from previous methods, which treat the hyperspectral images as multiple single channel images (reconstructing them separately) or as a whole, we divide the whole hyperspectral image into some groups. In this way, we can not only exploit the correlations among neighboring spectral bands of hyperspectral images, but also reduce the dimensionality of features of each group. Inspired by the success of the recently proposed residual network structure, which has achieved very good performance in the field of image restoration, we specifically design a SSB based on residual network structure. As shown in Fig. \ref{fig:N5new}, the proposed SSPSR network contains several branch networks and a global network. For each branch network and the global network, they first extract the shallow features and fed them to the SSPN, then upscale the outputs of SSPN with an intermediate upsampling factor. By cascading the parallel branch networks with the global network, we can super-resolve the input low-resolution hyperspectral image in a coarse-to-fine manner. In the following, we will give details of the branch network and global network, respectively.

\begin{figure*}[t]
  \centering
  \includegraphics[width=17.9cm]{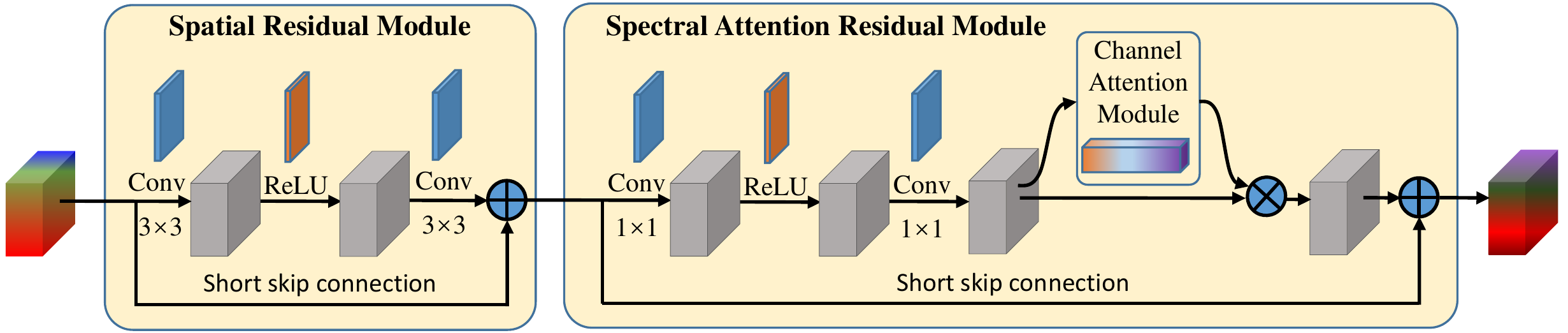}\\
  \caption{The network architecture of the spatial-spectral block (SSB), which consists of a spatial residual module and a spectral attention residual module. ``+'' and ``$\times$'' denote element-wise addition and element-wise multiplication, respectively.}\label{fig:SSPN}
\end{figure*}

\subsubsection{The Branch Network}
Specifically, the input low-resolution hyperspectral image $I_{LR}$ is firstly divided into $S$ groups, $I_{LR}=\{I_{LR}^{(1)}, I_{LR}^{(1)}, \cdots, I_{LR}^{(S)}\}$. It should be noted that, in our settings the neighboring groups may have overlaps. More details about the settings can be found at the experiment section. For each group $I_{LR}^{(s)}$, we directly apply one convolutional layer to obtain its shallow features $F_0^{(s)}$ as investigated in previous work \cite{lim2017enhanced, zhang2018image},
\begin{equation}\label{eq:F0}
F_{0}^{(s)} = H_{FE}(I_{LR}^{(s)}),
\end{equation}
where $H_{FE} (\cdot)$ denotes convolution operation, \emph{i.e.}, feature extraction layer. $F_{0}^{(s)}$ is then used for deep feature extraction with the proposed SSPN. Consequently, we can further have
\begin{equation}\label{eq:F0}
F_{SSPN}^{(s)} = H_{SSPN}(F_{0}^{(s)}),
\end{equation}
where $H_{SSPN} (\cdot)$ denotes the function of the proposed SSPN, which contains $R$ SSBs and we will present its details in the following.

The output of SSPN can be treated as the deep features of one grouped hyperspectral images. In order to alleviate the burden of the final super-resolution reconstruction, we adopt a strategy of progressive super-resolution reconstruction. Particularly, we add an upsampling module in the middle of the network (before feeding the output of branch SSPN to the global SSPN), which has proven to be a very effective technique, especially when the magnification is very large. Thus, by the upsampling module we obtain the upscaled feature maps,
\begin{equation}\label{eq:F0}
F_{UP}^{(s)} = H_{UP} (F_{SSPN}^{(s)}),
\end{equation}
where $H_{UP} (\cdot)$ and $F_{UP}^{(s)}$ denote an upsampling module and upscaled features respectively. In this paper, we leverage the \emph{PixelShuffle} \cite{Shi2016Real} operator to conduct the upsampling procedure.

Before feeding the upscaled features to the following global SSPN, we add one \emph{Conv} layer after each branch upsampling module to reduce the number of feature channels to the spectral number of each input group. Therefore, the output of the branch network will have the same channels as the input grouped hyperspectral images, and we call this layer as a ``reconstruction'' layer,
\begin{equation}\label{eq:rec}
F_{rec}^{(s)} = H_{rec} (F_{UP}^{(s)}),
\end{equation}
where $H_{rec} (\cdot)$ denotes the ``reconstruction'' layer (Here we use a lowercase term ``rec'' to represent a pseudo-reconstruction operation). By this \emph{Conv} layer, each branch can be seen as a super-resolution reconstruction subnetwork. Another purpose of designing this layer is to make the branch SSPN and global SSPN have the same network structure.

\subsubsection{The Global Network}
After extracting features from different groups with the branch networks, we concatenate them together from all branches (as shown in the ``concatenation operator'' of Fig. \ref{fig:N5new}), \emph{i.e.}, $F_{C}=[F_{rec}^{(1)},F_{rec}^{(2)},\cdots,F_{rec}^{(S)}]$. It should be noted that if the neighboring groups have overlaps, the integrated feature maps can be generated according to their original spectral band position and by averaging feature values in the overlapping bands. Similar to the local branch, before feeding the contacted features into the global SSPN, we apply one \emph{Conv} layer to extract the ``shallow features'',
\begin{equation}\label{eq:F0}
F_{G0} = H_{GFE}(F_{C}),
\end{equation}
where $H_{GFE}(\cdot)$ is similar to $H_{FE} (\cdot)$ and is used to extract corresponding ``shallow features'' of the input contacted features of all branch networks.

And then, we further feed $F_{G0}$ into the global SSPN, whose structure is the same as the local one,
\begin{equation}\label{eq:F0}
F_{GSSPN} = H_{GSSPN}(F_{G0}),
\end{equation}
where $H_{GSSPN}(\cdot)$ refers to the global version of $H_{SSPN}(\cdot)$. In this way, we extract the spatial-spectral features $F_{GSSPN}$ of the input hyperspectral images.

To upscale the obtained features to the target size, here we apply upsampling module once more (progressively reconstruction) to generate the upscaled spatial-spectral feature maps,
\begin{equation}\label{eq:F0}
F_{GUP} = H_{GUP} (F_{GSSPN}),
\end{equation}
where $H_{GUP}$ refers to the global version of $H_{UP}$.

The final super-resolved hyperspectral images can be then obtained via one reconstruction layer by feeding the upscaled spatial-spectral features and the upscaled input hyperspectral images,
\begin{equation}\label{eq:F0}
I_{SR} = H_{GREC}(F_{GUP}+H_{GFE2}(I_{LR} \uparrow)),
\end{equation}
where $I_{LR}\uparrow$ refers to the Bicubic upsampling version of the input low-resolution hyperspectral images, $H_{GFE2}(\cdot)$ is similar to $H_{GFE}(\cdot)$ and is used to extract shallow features of the input Bicubic upscaled hyperspectral images for residual learning, and $H_{GREC} (\cdot)$ is the reconstruction operation that has one $Conv$ layer. Here, ``$+H_{GFE}(I_{LR} \uparrow)$'' is referred to as the residual learning.

\subsection{Spatial-Spectral Prior Network (SSPN)}
Image super-resolution is a very ill-posed problem, which calls for additional prior (regularization) to constrain the reconstruction procedure. Traditional approaches all try to design sophisticated regularization terms such as total variation (TV), sparse, low-rank, by hand \cite{akhtar2015bayesian, dong2016hyperspectral, han2018self, zhang2018exploiting, veganzones2015hyperspectral}. Therefore, the performance of these algorithms is highly dependent on whether the designed prior can well characterize the observed data. As for the hyperspectral image super-resolution problem, it is crucial to effectively exploit the intrinsic properties of hyperspectral images, \emph{i.e.,} the non-local self-similarity in spatial and the high correlation across spectra. Previous manually designed constraints are insufficient for accurate hyperspectral image restoration.

In this paper, we advocate a spatial-spectral feature extraction network (SSPN) to exploit the spatial and spectral prior.
In particular, SSPN cascades $R$ spatial-spectral blocks (SSBs) and can be formulated as,
\begin{equation}\label{eq:F0}
F_{SSB_r}^{(s)} = H_{SSB_R}(H_{SSB_{R-1}}(\cdots H_{SSB_1}(F_{0}^{(s)}) \cdots )),
\end{equation}
where $H_{SSB_r}(\cdot)$ refers to the function of the $r$-th SSB, and $F_{SSB_{r-1}}^{(s)}$ is the input of the $r$-th SSB and $F_{SSB_r}^{(s)}$ is the extracted features. Noted that we use the notations from the local branch network to demonstrate the detailed design of the local SSPN, and the global SSPN is the same to the local one. %Without causing misunderstanding, in the following we will remove all superscript ``(s)'' for the sake of convenience.

To facilitate the prediction of the target, the long skip connection is further introduced in SSPN. This will lead to the direct passing of the low frequency features of the current features to the end, and let the current residual body pay more attention to the high frequency information. Therefore, the output of the SSPN can be obtained by
\begin{equation}\label{eq:F0}
F_{SSPN}^{(s)} = H_{SSB_R}(H_{SSB_{R-1}}(\cdots H_{SSB_1}(F_{0}^{(s)}) \cdots ))+F_0^{(s)}.
\end{equation}
Here,``$+F_{0}^{(s)}$'' is referred to as the residual learning (same as below). This residual in residual structure can enable fast as well as stable training.

In this paper, we specifically design the SSB to exploit the spatial-spectral information from the hyperspectral images. In particular, each SSB has two parts, \emph{i.e.} a spatial residual module and a spectral attention residual module. The architecture of SSB is illustrated in Fig. \ref{fig:SSPN}. For the spatial residual module, we leverage the standard residual block with 3$\times$3 convolutions to extract the spatial features,
\begin{equation}\label{eq:F0}
F_{Spa_r}^{(s)}=F_{SSB_{r-1}}^{(s)} + H_{SSB_r-Spa}(F_{SSB_{r-1}}^{(s)}),
\end{equation}
where $H_{SSB_r-Spa}(\cdot)$ refers to the function of the spatial residual module for the $r$-th SSB, and $F_{Spa_r}$ is the spatial feature for the $r$-th SSB. The standard residual block can well extract the spatial information of a hyperspectral image.

However, due to the strong correlation between the spectra of a hyperspectral image, standard residual convolutional networks cannot effectively extract the spectral dependencies. The spectral correlation, which is characterized by that there exists strong correlation among neighboring spectral bands of hyperspectral image, has been widely used for hyperspectral image reconstruction and analysis \cite{yokoya2011coupled, wycoff2013non}. To exploit this correlation, we can use all the spectral bands $\textbf{x}_1, \textbf{x}_2, ..., \textbf{x}_C$ to obtain the newly reconstructed spectral band $\textbf{x}_i^{'}$, \emph{i.e.}, $\textbf{x}_i^{'}=w_{i,1}\textbf{x}_1+w_{i,2}\textbf{x}_2+...+w_{i,C}\textbf{x}_C$. $\textbf{w}_i=[w_{i,1}, w_{i,2}, ..., w_{i,C}]$ are the linear combination (reconstruction) weights. If similar spectral bands share similar weights, the correlation information will be embedded in the reconstructed spectral band, thus exploiting the correlation among neighboring spectral bands of hyperspectral image. If we relax the weights to any learnable parameters, this will be equal to learning a set of weight vectors $\{\textbf{w}_i\}_i$, and thus obtaining a new representation of the hyperspectral image. Mathematically, this can be achieved by some 1$\times$1 filters (bottleneck layer), whose weights are $\{\textbf{w}_i\}_i$. By designing a spectral network with 1$\times$1 filters, we can expect to fully exploit the correlations between different spectral bands. It is worth noting that we further apply the ReLU layer to enhance its representation ability. Therefore, the structure of the SSB is designed as the combination of a spatial residual module and a spectral attention residual module as shown in Fig. \ref{fig:SSPN}. Thus, we have
\begin{equation}\label{eq:F0}
F_{{SSB}_{r}}^{(s)}=F_{Spa_{r}}^{(s)} + H_{SSB_r-Spc}(F_{Spa_{r}}^{(s)}),
\end{equation}
where $H_{SSB_r-Spc}(\cdot)$ denotes the spectral network of the $r$-th SSB.

To further improve the representation ability of spectral information as well as the entire network, we are inspired by Zhang \emph{et al}. \cite{zhang2018image} and introduce the channel attention mechanism to adaptively rescale each channel-wise feature by modeling the interdependencies across feature spectra. Specifically, a global average pooling layer is applied to the extracted feature maps of previous spectral network to obtain a global context embedding vector. And then, two thin fully connected layers with a simple gating mechanism (by sigmoid function) is applied to learn nonlinear interactions between spectra. Then we obtain the final channel scaling coefficient vector $T \in \mathbb{R}^{1\times 1 \times C}$, which is used to reweight the extracted feature maps. The output of the spectral attention residual module is simply computed by
\begin{equation}\label{eq:F0}
F_{{SSB}_{r}}^{(s)}=F_{Spa_{r}}^{(s)} + T H_{SSB_r-Spc}(F_{Spa_{r}}^{(s)}).
\end{equation}

\begin{table}
%\vspace{1mm}
\begin{center}
\caption{Average quantitative performance by different loss functions over four testing images of Chikusei dataset with respect to six PQIs when the upsampling factor is 4. }\label{tab:loss}
%\vspace{1.5mm}
%\scriptsize
%\setlength{\tabcolsep}{7.5pt}
\begin{tabular}{cccccc}
\hline
Losses  &   $l_2$ &   $l_1$ &    $l_1$+SSTV \\
    \hline
      \hline
CC $\uparrow$      &   0.9535   &    0.9560   &        0.9565   \\
SAM$\downarrow$      &   2.5152   &    2.3581   &       2.3527   \\
RMSE$\downarrow$     &   0.0117   &    0.0115   &        0.0114   \\
ERGAS$\downarrow$    &   5.1304   &    4.9903   &        4.9313   \\
PSNR$\uparrow$     &   40.0703  &    40.3515  &       40.3612   \\
SSIM$\uparrow$     &   0.9401   &    0.9437   &      0.9441   \\
    \hline
\end{tabular}
\normalsize
\end{center}
\end{table}

\subsection{Loss Function}
In order to measure the super-resolution performance, several cost functions have been investigated to make the super-resolution results approximate to ground truth high-resolution images. In the current literature, $l_2$, $l_1$, perceptual, and adversarial losses are the most commonly used loss functions. When compared with perceptual and adversarial losses, which may restore details that do not exist in the original images and is undesirable in remote sensing field, $l_2$ and $l_1$ losses are more credible. As for $l_2$ loss, it encourages finding pixel-wise averages of plausible solutions which are typically overly-smooth. Due to that $l_1$ loss can effectively penalize small errors and maintain better convergence throughout the training phase, we adopt $l_1$ loss to measure the reconstruction accuracy of the network. Specifically, the $l_1$ loss is defined by mean absolute error (MAE) between all the reconstructed images and the ground truth:
\begin{equation}\label{eq:F0}
\mathcal{L}_1(\Theta)  =  \frac{1}{N}  \sum_{n=1}^N   \left\| {I_{HR}^n  -  H_{Net}(I_{LR}^n)} \right\|_1,
\end{equation}
where $H_{Net}(I_{LR}^n)$ and $I_{HR}^n$ are the $n$-th reconstructed high-resolution hyperspectral image and ground truth hyperspectral image, respectively. $N$ denotes the number of images in one training batch, and $\Theta$ refers the parameter set of our network.

However, above-mentioned loss is primarily designed for general image restoration tasks. Although they can well preserve the spatial information of the super-resolution results, the reconstructed spectral information may be distorted due to the ignorance of the correlations among spectral features. In order to simultaneously ensure the spatial and spectral credibility of the reconstruction results, we introduce the spatial-spectral total variation (SSTV) \cite{aggarwal2016hyperspectral}. It extends the conventional total variation model and accounts for both the spatial and the spectral correlation. In this paper, we add the SSTV to the $l_1$ loss to impose spatial and spectral smoothness simultaneously,
\begin{equation}\label{eq:F0}
\mathcal{L}_{SSTV}(\Theta)  =  \frac{1}{N}  \sum_{n=1}^N   (\left\| {\nabla_h I_{SR}^n} \right\|_1+\left\| {\nabla_w I_{SR}^n} \right\|_1+\left\| {\nabla_c I_{SR}^n} \right\|_1),
\end{equation}
where $\nabla_h$, $\nabla_w$, and $\nabla_c$ are functions to compute the horizontal, vertical, and spectral gradient of $I_{SR}^n$.

In summary, the final objective loss for the proposed model is a weighted sum of the two losses:
\begin{equation}\label{eq:F0}
\mathcal{L}_{total}(\Theta) = \mathcal{L}_{1} + \alpha \mathcal{L}_{SSTV},
\end{equation}
where $\alpha$ is used to balance the contributions of different losses. In our experiments, we set it as a constant, $\alpha=1e-3$.

In Table \ref{tab:loss}, we report the reconstruction results (in terms of objective measurements) when using different losses (more details regarding the experimental settings can be found at the experiment section). Clearly, $l_1$ loss is much more suitable for our task, because it can effectively penalize small errors and maintain better convergence throughout the training phase. By introducing the SSTV constraint, slightly better results can be achieved.

\subsection{Implementation Details}
We use Pytorch libraries\footnote{\url{https://pytorch.org}} to implement and train the proposed SSPSR network. We train different models to super-resolve the hyperspectral images for scale factors 4 and 8 with random initialization. We use the ADAM optimizer \cite{kingma2014adam} with an initial learning rate of 1e-4 which decays by a factor of 10 when it reaches 30 epochs. In our experiments, we find it will take 40 epochs to achieve a stable performance. The models are trained with a batch size of 32. As in many previous work, we also apply the Bicubic interpolation to downsample the high-resolution hyperspectral images to obtain the corresponding low-resolution hyperspectral images.

Unless otherwise specified, in the following experiments we set the spectral band number ($p$) of each group to 8 and the overlap ($o$) between neighboring groups to 2. To efficiently process the ``edge'' spectral bands, we adopt a so called ``fallback'' dividing strategy. When the last group has less than $p$ spectral bands, we select the last $p$ bands as the last group. Therefore, the number of groups can be obtained by the following equation,
\begin{equation}\label{eq:groups}
S = ceil\left( \frac{C-o}{p-o} \right),
\end{equation}
where $ceil(\cdot)$ is the function that rounds the elements of to the nearest integers towards infinity. In the SSPN, the number of spatial-spectral blocks ($R$) is set to 3.
We set the size of all \emph{Conv} layers to 3$\times$3 except for that in the spectral residual modules, where the kernel size is set to 1$\times$1. To ensure that the size of the feature map is not changed, the zero-padding strategy is applied for these \emph{Conv} layers with kernel size 3$\times$3. The \emph{Conv} layers in shallow feature extraction and SSPN have $C=256$ filters, except for that in the channel-downscaling, \emph{i.e.}, the reconstruction network after the upscaled features at the branch networks (please refer to Eq. (\ref{eq:rec})).

\setlength{\tabcolsep}{2.50pt}
\begin{table}
\vspace{1mm}
\begin{center}
\caption{Ablation study. Quantitative comparisons among some other variants of the proposed SSPSR method over four testing images of Chikusei dataset with respect to six PQIs.}\label{tab:Ablation}
\vspace{1.5mm}
%\scriptsize
%\setlength{\tabcolsep}{7.5pt}
\begin{tabular}{cccccccccc}
  \hline
Models	&	$d$	&	CC$\uparrow$	&	SAM$\downarrow$	&	RMSE$\downarrow$	&	ERGAS$\downarrow$	&	PSNR$\uparrow$	&	SSIM$\uparrow$	\\
\hline
\hline
Our 	&	4	&	0.9565	&	2.3527	&	0.0114	&	4.9313	&	40.3612	&	0.9441	\\
Our - w/o GS	&	4	&	0.9548	&	2.4048	&	0.0116	&	5.0399	&	40.1901	&	0.9424	\\
Our - w/o PU	&	4	&	0.9520	&	2.5239	&	0.0119	&	5.2329	&	39.9185	&	0.9388	\\
Our - w/o PS	&	4	&	0.9537	&	2.4152	&	0.0118	&	5.0991	&	40.0712	&	0.9410	\\
Our - w/o SA	&	4	&	0.9563	&	2.3597	&	0.0115	&	4.9443	&	40.3408	&	0.9438	\\
\hline
\hline
Our 	&	8	&	0.8766	&	4.0127	&	0.0191	&	8.3355	&	35.8368	&	0.8538	\\
Our - w/o GS	&	8	&	0.8622	&	4.5121	&	0.0199	&	8.8459	&	35.3857	&	0.8427	\\
Our - w/o PU	&	8	&	0.8585	&	4.5542	&	0.0202	&	9.0285	&	35.2489	&	0.8358	\\
Our - w/o PS	&	8	&	0.8732	&	4.0587	&	0.0194	&	8.4621	&	35.7074	&	0.8522	\\
Our - w/o SA	&	8	&	0.8760	&	4.0198	&	0.0192	&	8.3650	&	35.8144	&	0.8538	\\
  \hline
\end{tabular}
\normalsize
\end{center}
\text{\footnotesize{SG: Grouping Strategy, PU: Progressive Upsampling}}
\text{\footnotesize{PS: Parameter Sharing, SA: Spectral Attention}}
\end{table}

\begin{table*}
%\vspace{1mm}
\begin{center}
\caption{The performance of some typical setting for the spectral band numbers of each group and overlaps between neighboring groups when using the grouping strategy of the proposed SSPSR method.}\label{tab:GS}
%\vspace{1.5mm}
\footnotesize
\begin{tabular}{ccc|cc|ccccccc}

  \hline
bands ($p$)  &   overlaps ($o$)  &   groups ($S$)     &   params$\times 10^6$    &   FLOPs$\times 10^9$   &	CC$\uparrow$	&	SAM$\downarrow$	&	RMSE$\downarrow$	&	ERGAS$\downarrow$	&	PSNR$\uparrow$	&	SSIM$\uparrow$	\\
\hline
\hline
128  &   0  & 1 &   14.12 &   11.16   &   0.9548	&	2.4048	&	0.0116	&	5.0399	&	40.1901	&	0.9424	\\
\hline
1    &   0  & 128 &   13.53 &   215.87   &   0.9558	&	2.3456	&	0.0116	&	4.9609	&	40.2757	&	0.9432	\\
\hline
8    &   0  & 16 &   13.56 &   35.34   &   0.9562	&	2.3670	&	0.0115	&	4.9540	&	40.3286	&	0.9437	\\
8    &   2  & 21 &   13.56 &   43.51   &	  0.9565	&	2.3527	&	0.0114	&	4.9313	&	40.3612	&	0.9441	\\
8    &   4  & 31 &   13.56 &   59.87   &   0.9567	&	2.3520	&	0.0114	&	4.9251	&	40.3759	&	0.9443	\\
8    &   6  & 61 &   13.56 &   108.94   &   0.9568	&	2.3512	&	0.0113	&	4.9205	&	40.3801	&	0.9445	\\
\hline
\end{tabular}
%\normalsize
\end{center}
\end{table*}

\section{Experiments and Results}
\label{sec:Experiments}
In this section, we present a detailed analysis and evaluation of our approach on three public hyperspectral image datasets, which include two remote sensing hyperspectral image datasets, \emph{i.e.}, Chikusei dataset \cite{NYokoya2016}\footnote{\url{https://www.sal.t.u-tokyo.ac.jp/hyperdata/}} and Pavia Center dataset\footnote{\url{http://www.ehu.eus/ccwintco/index.php?title=Hyperspectral_Remote_Sensing_Scenes}}, and one nature hyperspectral image dataset, \emph{i.e.}, CAVE dataset \cite{YasumaGeneralized}\footnote{\url{https://www.cs.columbia.edu/CAVE/databases/multispectral/}}. We compare the proposed method with eight comparison methods, including four state-of-the-art deep single gray/RGB image super-resolution methods, VDSR \cite{kim2016accurate}, EDSR \cite{lim2017enhanced}, RCAN \cite{zhang2018image}, and SAN \cite{dai2019second}, and four representative and most relevant deep single hyperspectral image super-resolution methods, TLCNN \cite{yuan2017hyperspectral}, 3DCNN \cite{mei2017hyperspectral}, GDRRN \cite{li2018single}, and DeepPrior \cite{sidorov2019deep}. We carefully adjust hyperparameters of these comparison methods to achieve their best performance. Bicubic interpolation is introduced as the baseline.

\textbf{Evaluation measures}. Six widely used quantitative picture quality indices (PQIs) are employed to evaluate the performance of our method, including cross correlation (CC) \cite{loncan2015hyperspectral}, spectral angle mapper (SAM) \cite{yuhas1992discrimination}, root mean squared error (RMSE), erreur relative globale adimensionnelle de synthese (ERGAS) \cite{wald2002data}, peak signal-to-noise ratio (PSNR), and structure similarity (SSIM) \cite{wang2004image}. For PSNR and SSIM of the reconstructed hyperspectral images, we report their mean values of all spectral bands. CC, SAM, and ERGAS are three widely adopted quality indices in HS fusion task, while the remaining three indices are commonly used quantitative image restoration quality indices. The best values for these indices are 1, 0, 0, 0, $+\propto$, and 1, respectively.

\begin{table}
%\vspace{1mm}
\begin{center}
\caption{Average quantitative comparisons of ten different approaches over four testing images from Chikusei dataset with respect to six PQIs. }\label{tab:Chikusei}
%\vspace{1.5mm}
%\footnotesize
%\setlength{\tabcolsep}{7.5pt}
\begin{tabular}{cccccccc}

  \hline
	&	$d$	&	CC$\uparrow$	&	SAM$\downarrow$	&	RMSE$\downarrow$	&	ERGAS$\downarrow$	&	PSNR$\uparrow$	&	SSIM$\uparrow$	\\
\hline
\hline
Bicubic	&	4	&	0.9212 	&	3.4040 	&	0.0156 	&	6.7564 	&	37.6377 	&	0.8949 	\\
\hline
VDSR \cite{kim2016accurate}	&	4	&	0.9227 	&	3.6642 	&	0.0148 	&	6.8708 	&	37.7755 	&	0.9065 	\\
EDSR \cite{lim2017enhanced}	&	4	&	0.9510 	&	2.5580 	&	0.0121 	&	5.3708 	&	39.8289 	&	0.9354 	\\
RCAN \cite{zhang2018image}	&	4	&	\underline{0.9518}	&	2.5397 	&	\underline{0.0120}	&	\underline{5.3205}	&	\underline{39.9041}	&	\underline{0.9359}	\\
SAN \cite{dai2019second}	&	4	&	0.9514 	&	2.5547 	&	\underline{0.0120}	&	5.3349 	&	39.8671 	&	0.9357 	\\
\hline
TLCNN \cite{yuan2017hyperspectral}	&	4	&	0.9196 	&	3.8573 	&	0.0150 	&	6.7522 	&	37.7251 	&	0.9008 	\\
3DCNN \cite{mei2017hyperspectral}	&	4	&	0.9355 	&	3.1174 	&	0.0140 	&	6.0026 	&	38.6091 	&	0.9127 	\\
GDRRN \cite{li2018single}	&	4	&	0.9369 	&	\underline{2.500}	&	0.0137 	&	5.9540 	&	38.7198 	&	0.9193 	\\
DeepPrior \cite{sidorov2019deep}	&	4	&	0.9293 	&	3.5590 	&	0.0147 	&	6.2096 	&	38.1923 	&	0.9010 	\\
\hline
SSPSR	&	4	&	\textbf{0.9565}	&	\textbf{2.3527}	&	\textbf{0.0114}	&	\textbf{4.9894}	&	\textbf{40.3612}	&	\textbf{0.9413}	\\
    \hline
    \hline
Bicubic	&	8	&	0.8314 	&	5.0436 	&	0.0224 	&	4.8488 	&	34.5049 	&	0.8228 	\\
\hline
VDSR \cite{kim2016accurate}	&	8	&	0.8344 	&	5.1778 	&	0.0216 	&	4.9052 	&	34.5661 	&	0.8305 	\\
EDSR \cite{lim2017enhanced}	&	8	&	0.8636 	&	4.4205 	&	0.0201 	&	\underline{4.5091}	&	35.4217 	&	0.8501 	\\
RCAN \cite{zhang2018image}	&	8	&	\underline{0.8665}	&	4.3757 	&	\underline{0.0198}	&	4.5229 	&	\underline{35.5044}	&	\underline{0.8531}	\\
SAN \cite{dai2019second}	&	8	&	0.8664 	&	4.3922 	&	\underline{0.0198}	&	4.5170 	&	35.5018 	&	0.8527 	\\
\hline
TLCNN \cite{yuan2017hyperspectral}	&	8	&	0.8249 	&	5.3041 	&	0.0224 	&	4.8843 	&	34.3488 	&	0.8215 	\\
3DCNN \cite{mei2017hyperspectral}	&	8	&	0.8428 	&	4.8432 	&	0.0215 	&	4.5964 	&	34.8375 	&	0.8313 	\\
GDRRN \cite{li2018single}	&	8	&	0.8421 	&	\underline{4.3160}	&	0.0214 	&	4.5879 	&	34.8153 	&	0.8357 	\\
DeepPrior \cite{sidorov2019deep}	&	8	&	0.8366 	&	5.3386 	&	0.0219 	&	4.6789 	&	34.6692 	&	0.8126 	\\
\hline
SSPSR	&	8	&	\textbf{0.8766}	&	\textbf{4.0127}	&	\textbf{0.0191}	&	\textbf{4.3120}	&	\textbf{35.8368}	&	\textbf{0.8624}	\\
  \hline
\end{tabular}
\normalsize
\end{center}
\end{table}

\begin{table}
%\vspace{1mm}
\begin{center}
\caption{Average quantitative comparisons of ten different approaches over four testing images from Pavia Centre dataset with respect to six PQIs. }\label{tab:Pavia}
%\vspace{1.5mm}
%\footnotesize
%\setlength{\tabcolsep}{7.5pt}
\begin{tabular}{cccccccc}

  \hline
	&	$d$	&	CC$\uparrow$	&	SAM$\downarrow$	&	RMSE$\downarrow$	&	ERGAS$\downarrow$	&	PSNR$\uparrow$	&	SSIM$\uparrow$	\\
    \hline
\hline
Bicubic	&	4	&	0.8594 	&	6.1399 	&	0.0437 	&	6.8814 	&	27.5874 	&	0.6961 	\\
\hline
VDSR \cite{kim2016accurate}	&	4	&	0.8659 	&	6.7004 	&	0.0419 	&	6.6991 	&	27.8821 	&	0.7242 	\\
EDSR \cite{lim2017enhanced}	&	4	&	0.8922 	&	5.8657 	&	0.0379 	&	6.0199 	&	28.7981 	&	0.7722 	\\
RCAN \cite{zhang2018image}	&	4	&	0.8917 	&	5.9785 	&	0.0376 	&	6.0485 	&	28.8165 	&	0.7719 	\\
SAN \cite{dai2019second}	&	4	&	\underline{0.8927}	&	5.9590 	&	\underline{0.0374}	&	\underline{5.9903}	&	\underline{28.8554}	&	\underline{0.7740}	\\
\hline
TLCNN \cite{yuan2017hyperspectral}	&	4	&	0.8563 	&	6.9013 	&	0.0431 	&	6.9139 	&	27.6682 	&	0.7141 	\\
3DCNN \cite{mei2017hyperspectral}	&	4	&	0.8813 	&	5.8669 	&	0.0396 	&	6.2665 	&	28.4114 	&	0.7501 	\\
GDRRN \cite{li2018single}	&	4	&	0.8829 	&	\underline{5.4750}	&	0.0393 	&	6.2264 	&	28.4726 	&	0.7530 	\\
DeepPrior \cite{sidorov2019deep}	&	4	&	0.8723 	&	6.2665 	&	0.0410 	&	6.4845 	&	28.1061 	&	0.7365 	\\
\hline
SSPSR	&	4	&	\textbf{0.9003}	&	\textbf{5.4612}	&	\textbf{0.0362}	&	\textbf{5.8014}	&	\textbf{29.1581}	&	\textbf{0.7903}	\\
    \hline
    \hline
Bicubic	&	8	&	0.6969 	&	7.8478 	&	0.0630 	&	4.8280 	&	24.5972 	&	0.4725 	\\
\hline
VDSR \cite{kim2016accurate}	&	8	&	0.7116 	&	8.0769 	&	0.0611 	&	4.6851 	&	24.8483 	&	0.5017 	\\
EDSR \cite{lim2017enhanced}	&	8	&	\underline{0.7215}	&	7.8594 	&	\underline{0.05983}	&	4.6359 	&	\underline{25.0041}	&	\underline{0.5130}	\\
RCAN \cite{zhang2018image}	&	8	&	0.7152 	&	7.9992 	&	0.0604 	&	4.6930 	&	24.9183 	&	0.5086 	\\
SAN \cite{dai2019second}	&	8	&	0.7104 	&	8.0371 	&	0.0609 	&	4.7646 	&	24.8485 	&	0.5054 	\\
\hline
TLCNN \cite{yuan2017hyperspectral}	&	8	&	0.6880 	&	8.3843 	&	0.0633 	&	4.9143 	&	24.5215 	&	0.4790 	\\
3DCNN \cite{mei2017hyperspectral}	&	8	&	0.7163 	&	7.6878 	&	0.0605 	&	4.6469 	&	24.9336 	&	0.5038 	\\
GDRRN \cite{li2018single}	&	8	&	0.7111 	&	\underline{7.3531}	&	0.0607 	&	\underline{4.6220}	&	24.8648 	&	0.5014 	\\
DeepPrior \cite{sidorov2019deep}	&	8	&	0.7007 	&	7.9281 	&	0.0618 	&	4.7366 	&	24.7252 	&	0.4963 	\\
\hline
SSPSR	&	8	&	\textbf{0.7359}	&	\textbf{7.3312}	&	\textbf{0.0586}	&	\textbf{4.5266}	&	\textbf{25.1985}	&	\textbf{0.5365}	\\
  \hline
\end{tabular}
\normalsize
\end{center}
\end{table}

\subsection{Ablation Studies}
The proposed SSPSR method contains four main components including Grouping Strategy (GS), Progressive Upsampling (PU), Parameter Sharing (PS), and Spectral Attention (SA). In order to validate the effectiveness of these components, we modify our model and compare their variants. We use the training images from Chikusei dataset as a training set, and evaluate the super-resolution performance (in terms of average objective results) on the four testing images from Chikusei dataset (more details regarding the experimental settings on Chikusei dataset can be found in the following subsection). Table \ref{tab:Ablation} tabulates the four variants of the proposed method, in which $d$ denotes the upsampling scale. In the following, we will give the detailed analysis about them.

\textbf{Grouping Strategy (GS)}. To effectively exploit the correlation among neighboring spectral bands of hyperspectral image and reduce the parameters of the model, we design a grouping strategy to divide the input hyperspectral image into some overlap groups. In order to verify the effectiveness of this strategy, we remove the grouping strategy and treat them as one group. As shown in Table \ref{tab:Ablation}, ``Our - w/o GS'', where the grouping strategy is discarded, is getting worse. The grouping strategy leads to a considerable performance improvement, \emph{e.g.},+0.17 dB for $\times$4 and +0.45 dB for $\times$8. As for other objective indicators, the gains are also considerable.

In addition to above with/without GS comparisons, we also report the number of parameters and FLOPs as well as the six PQIs of our method under some typical setting for the spectral band numbers ($p$) of each group and overlaps ($o$) between neighboring groups. The group number $S$ is calculated by Eq. (\ref{eq:groups}). As shown in Table \ref{tab:Ablation}, when $p=128$ and $o=0$, our method considers all the spectral bands as a whole group ($S=1$) and there is no grouping strategy, \emph{i.e.}, the case of ``Our - w/o GS''. When $p=1$, $o=0$, and $S=128$ our method will treat each spectral band as a group and this can be seen as a special case, \emph{i.e.}, the band-wise grouping. From the results, we can see that regardless of whether we treat all spectra as a whole or treat them separately, their performance cannot be compared with our proposed grouping strategy. When comparing the two schemes, the band-wise one obtained better performance due to the combination of grouping and parameter sharing. However, it will also greatly increase the computational overhead (please refer to the FLOPs). Because the more branches of the model, the more calculations are required.

\begin{figure*}
  \centering
  \includegraphics[width=16cm]{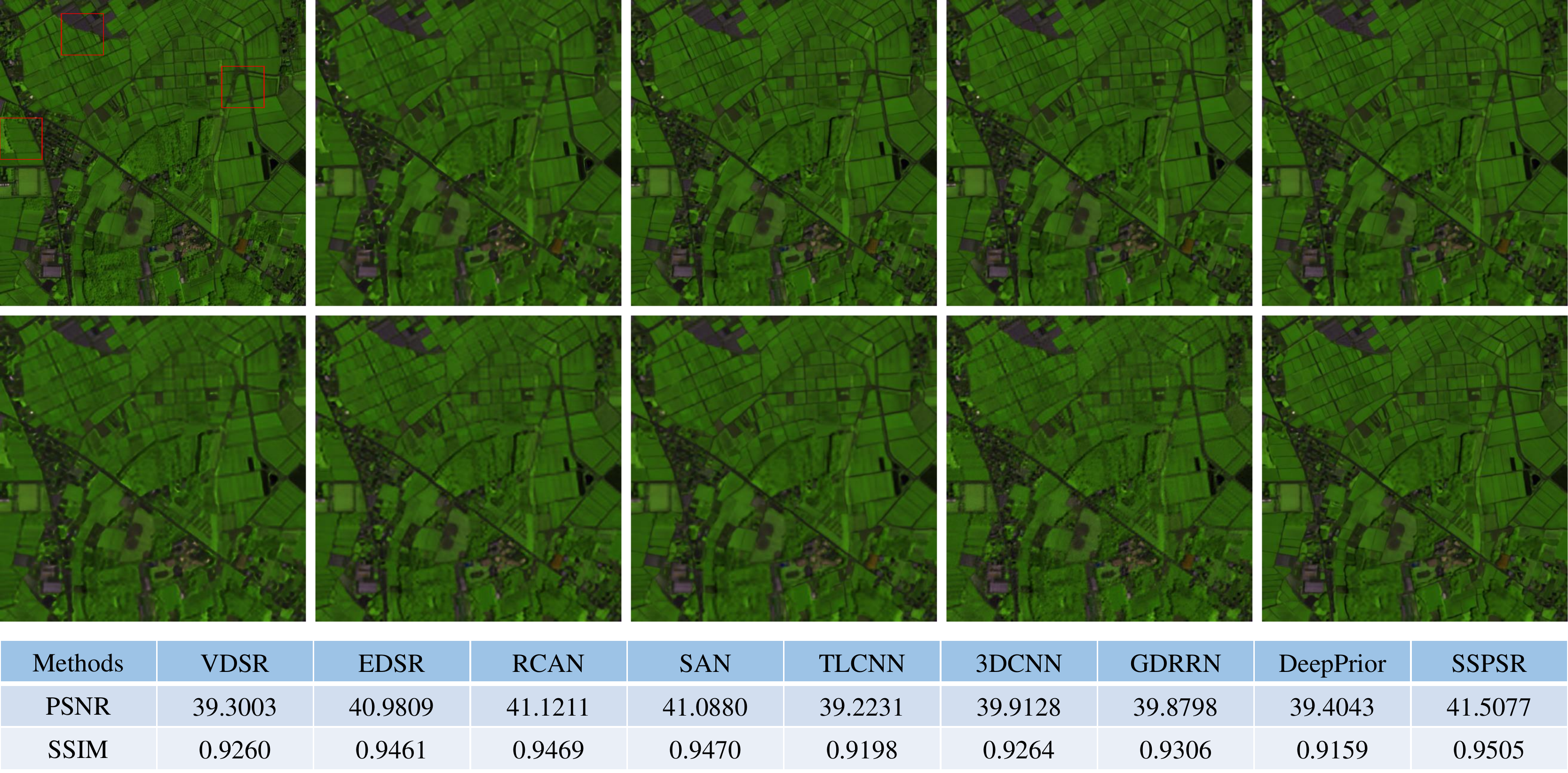}\\
  \caption{Reconstructed composite images of one test hyperspectral image in Chikusei dataset with spectral bands 70-100-36 as R-G-B when the upsampling factor is $d$ = 4. From left to right, top to down, they are the ground truth, results of VDSR \cite{kim2016accurate}, EDSR \cite{lim2017enhanced}, RCAN \cite{zhang2018image}, SAN \cite{dai2019second}, TLCNN \cite{yuan2017hyperspectral}, 3DCNN \cite{mei2017hyperspectral}, GDRRN \cite{li2018single}, DeepPrior \cite{sidorov2019deep}, and the proposed SSPSR method. The bottom table shows the PSNR (dB) and SSIM results of the reconstructed RGB composite image of different methods.}\label{fig:japanx4}
\end{figure*}

We also report the performance of the proposed methods with different settings for the overlaps between neighboring groups, \emph{i.e.}, $p=8$ and $o=0, 2, 4, 6$. With the increase of overlap (from $o=0$ to $o=6$), the performance of our method will be gradually improved, but the calculation amount of the model is also constantly expanding. It is worth noting that because we adopt a strategy of parameter sharing, when we fix the spectral band number $p$ and change the overlap $o$, the parameters of the model are the same. In order to achieve a balance among the number of parameters and FLOPs and the objective results, in this paper, we set the $p$ and $o$ to 8 and 2, respectively.

\textbf{Progressive Upsampling (PU)}. To learn the end-to-end relationship between low-resolution input and high-resolution output, there are two commonly used upsampling frameworks, pre-upsampling super-resolution and post-upsampling super-resolution. They either increase the parameters of the network or increase the difficulty of training.
Inspired by Laplacian pyramid super-resolution network \cite{lai2017deep}, we leverage a progressive upsampling super-resolution framework. In this way, it decomposes a difficult task into some easy tasks, thus not only greatly reducing the learning difficulty but also obtaining better performance. In Table \ref{tab:Ablation}, we report the performance of the proposed SSPSR method without the PU strategy, \emph{i.e.}, ``Our - w/o PU''. We remove the upsampling module in the branch networks and obtain the variant of our method.
We can see that our method with PU achieves better performance on all the six indices, including the spatial reconstruction fidelity (\emph{e.g.}, RMSE, PSNR and SSIM) and the spectral consistency (CC, SAM, and ERGAS). Especially when the upsampling factor is large, this strategy appears to be paramount. For example, the improvement of CC and PSNR of $\times$8 is greater than that of $\times$4, \emph{e.g.}, +0.045 and +0.45 dB for $\times$4, and +0.181 and +0.58 dB for $\times$8.

\begin{figure*}
  \centering
  \includegraphics[width=16cm]{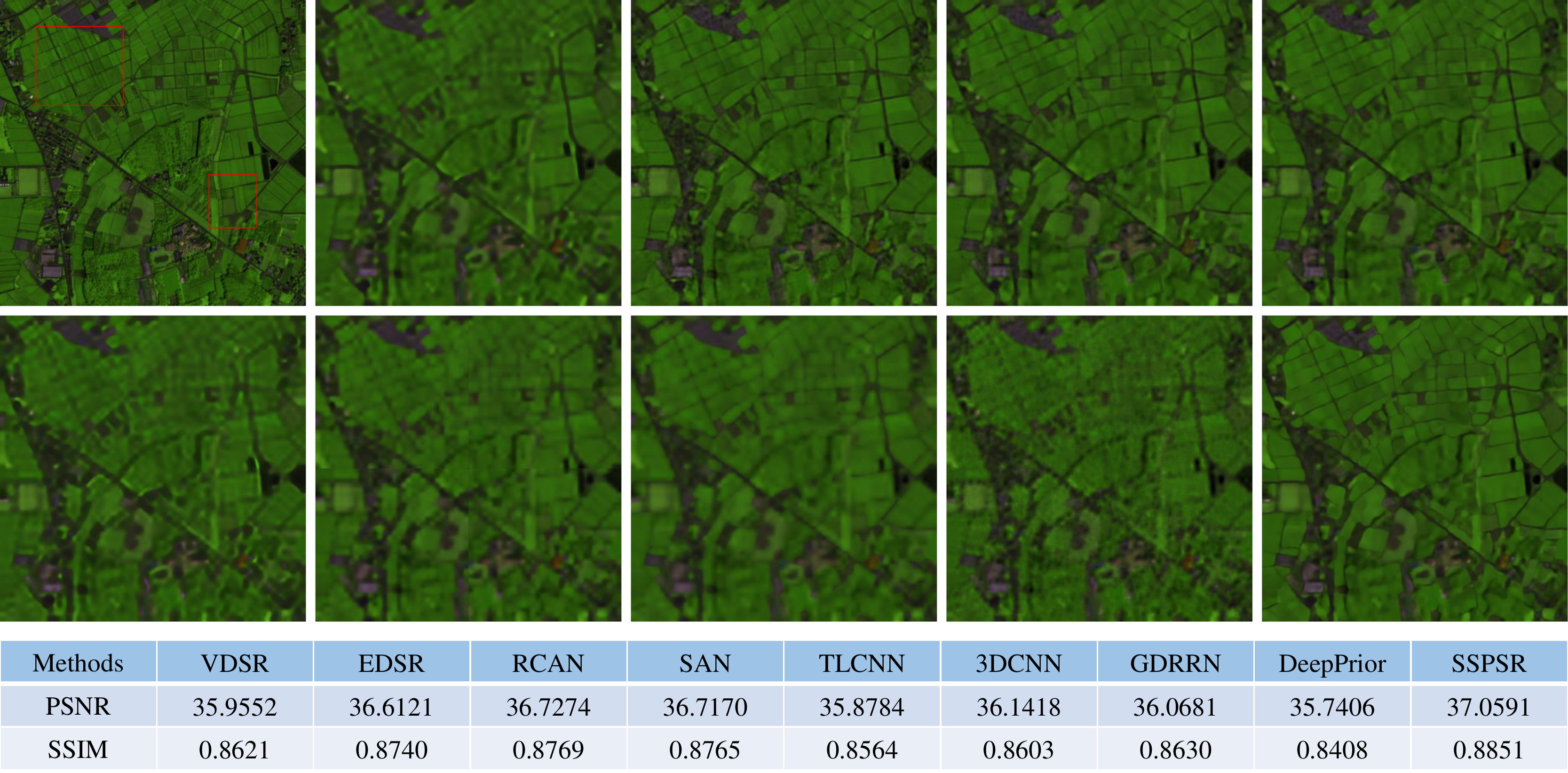}\\
  \caption{Reconstructed composite images of one test hyperspectral image in Chikusei dataset with spectral bands 70-100-36 as R-G-B when the upsampling factor is $d$ = 8. From left to right, top to down, they are the ground truth, results of VDSR \cite{kim2016accurate}, EDSR \cite{lim2017enhanced}, RCAN \cite{zhang2018image}, SAN \cite{dai2019second}, TLCNN \cite{yuan2017hyperspectral}, 3DCNN \cite{mei2017hyperspectral}, GDRRN \cite{li2018single}, DeepPrior \cite{sidorov2019deep}, and the proposed SSPSR method. The bottom table shows the PSNR (dB) and SSIM results of the reconstructed RGB composite image of different methods.}\label{fig:japanx8}
\end{figure*}

\textbf{Parameter Sharing (PS)}. In the proposed SSPSR method, in order to make the training process more efficient, we share the network parameters of each branch across all groups. In Table \ref{tab:Ablation}, we tabulates the comparison results of the proposed SSPSR method with and without parameter sharing strategy. Obviously, by parameter sharing, we have greatly reduced the computational complexity of the model. Although parameter sharing strategy reduces the parameters of the model, it does not weaken the representation ability of the model. Through the parameter sharing strategy\footnote{Since the network parameters are mainly dominated by module of SSPN, we can deduce that the parameter ratio between the models with and without parameter sharing is $\frac{2}{S+1}$.}, we can make full use of the training samples provided by different branches (training ``more'' data with only one branch network parameters), so that we get a more stable model. From the results, we can see that the overall performance of the parameter sharing strategy is even better than the parametric unsharing method on all six PQIs under $d=4$ and $d=8$.

\textbf{Spectral Attention (SA)}. To exploit the spatial-spectral prior, we apply the bottleneck network (with 1$\times$1 filters) to extract the correlations among neighboring spectral bands of hyperspectral image. In addition, the attention module is also introduced to model the interdependencies between the spectra of the hyperspectral data. To verify the effectiveness of the SA module, we compare the performance of with and without SA module. As shown in Table \ref{tab:Ablation}, with the SA mechanism, our method has achieves a slight performance gain compared to ``Our - w/o SA'' that without SA mechanism. By adding the SA module, although the improvement of each objective index is relatively small, the improvement of spectral confidence (\emph{i.e.}, SAM) is more obvious than that of spatial reconstruction confidence (\emph{i.e.}, PSNR), 2.2\% vs. 0.43\% for $d=4$ and 11\% vs. 1.3\% for $d=4$. This proves that the introduction of SA will be more conducive to the representation of spectral features.

\subsection{Results on Chikusei Dataset}
The Chikusei dataset is taken by Headwall Hyperspec-VNIR-C imaging sensor, and it is an urban area in Chikusei, Ibaraki, Japan, taken on 29 July 2014. It has 128 spectral bands in the spectral range from 363 nm to 1018 nm and 2517$\times$2335 pixels in total.

Due to missing information on the edge, we first crop the center region of the image to obtain a subimage with 2304$\times$2048$\times$128 pixels, which is further divided into training and test data. Specifically, the top region of this image are extracted to form the testing data, which has four non-overlap hyperspectral images with 512$\times$512$\times$128 pixels. Besides, from the remaining region of the subimage, we extract overlap patches as reference high-resolution hyperspectral images for training (10\% of the training data is included as a validation set). When the upsampling factor $d$ is 4, we let the extracted patches as 64$\times$64 pixels (with 32 pixels overlap); when the upsampling factor $d$ is 8, we let the extracted patches as 128$\times$128 pixels (with 64 pixels overlap). Here we use different block sizes for different factors mainly because of the following considerations: if the factor is large and the patch size is small, the input information is very limited and this will hinder the training of the network. Therefore, we use a big patch size for the large factor. Note that the low-resolution hyperspectral images is generated by Bicubic downsampling (the Matlab function \emph{imresize}) the ground truth with a factor of 4 or 8.

Table \ref{tab:Chikusei} reports the average objective performance over four testing images of all comparison algorithms, where bold represents the best result, underline denotes the second best. We can easily observe that the proposed SSPSR method significantly outperforms other algorithms with respect to all objective evaluation indexes. The average PSNR value of our method is more than 0.30 dB higher than that of the second best method. As a two-step method (first super-resolves the hyperspectral images and then conduct decomposition), TLCNN \cite{yuan2017hyperspectral} can well reconstruct the target hyperspectral images. Similar to our method, GDRRN \cite{li2018single} also takes a group strategy, and thus can well exploit the spectral information (it achieves the second best results in term of SAM). DeepPrior \cite{sidorov2019deep} is a very novel method, however, it takes much time to adjust the results and there is no superior strategy to determine when to stop iteration. RCAN \cite{zhang2018image} and SAN \cite{dai2019second} receive the similar results and are slight better than EDSR \cite{lim2017enhanced}. This may be due to the fact that the former two consider the channel attention, and thus can well capture the spectral features of the hyperspectral data.

Fig. \ref{fig:japanx4} and Fig. \ref{fig:japanx8} show the reconstructed composite images of one test hyperspectral image in Chikusei dataset of different comparison methods with upsampling factors $d$ = 4 and $d$ = 8, respectively. We can also easily observe that the proposed SSPSR method performs better than other algorithms, in the better recovery of both finer-grained textures and coarser-grained structures (please refer to the regions marked with red boxes). At the bottom of these visual comparison results, we also report their PSNR and SSIM values of the reconstructed composite images. Our approach SSPSR still has considerable advantages.

\begin{figure*}[t]
  \centering
  \includegraphics[width=17cm]{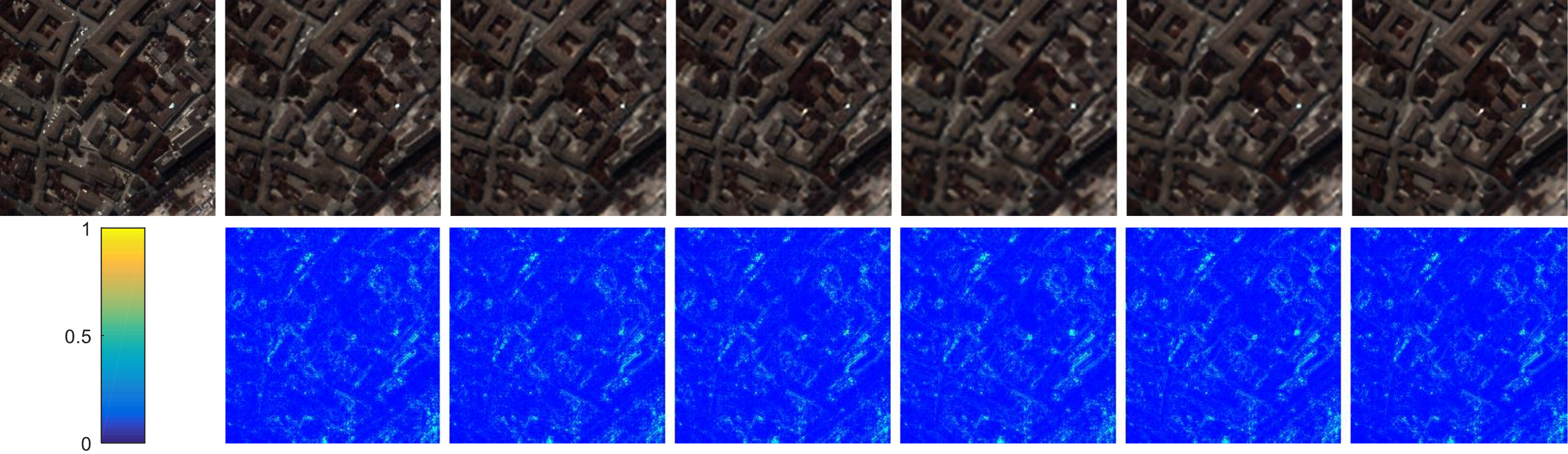}\\
  \caption{Reconstructed composite images (the first row) and the error maps (the second row)  of one test hyperspectral image in Pavia Center dataset with spectral bands 32-21-11 as R-G-B with upsampling factor $d$ = 4. From left to right, they are the ground truth, results of EDSR \cite{lim2017enhanced}, RCAN \cite{zhang2018image}, SAN \cite{dai2019second}, 3DCNN \cite{mei2017hyperspectral}, GDRRN \cite{li2018single}, and the proposed SSPSR method. The bottom images are the reconstruction error maps of the corresponding methods.}\label{fig:paviax4}
\end{figure*}

\begin{figure*}[t]
  \centering
  \includegraphics[width=17cm]{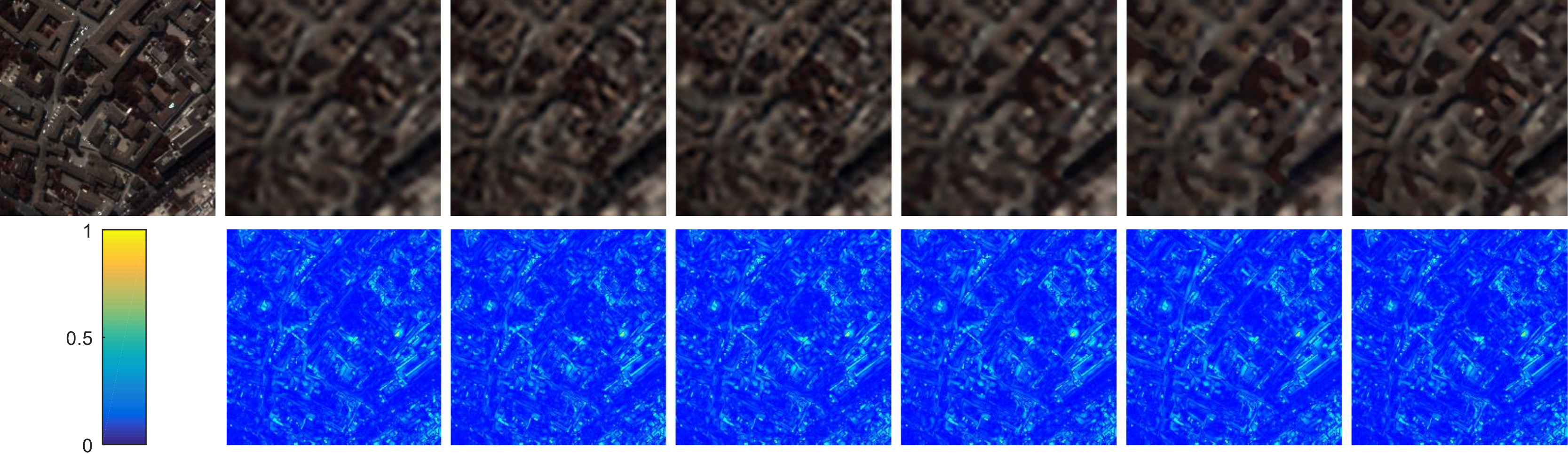}\\
  \caption{Reconstructed composite images (the first row) and the error maps (the second row)  of one test hyperspectral image in Pavia Center dataset with spectral bands 32-21-11 as R-G-B with upsampling factor $d$ = 8. From left to right, they are the ground truth, results of EDSR \cite{lim2017enhanced}, RCAN \cite{zhang2018image}, SAN \cite{dai2019second}, 3DCNN \cite{mei2017hyperspectral}, GDRRN \cite{li2018single}, and the proposed SSPSR method. The bottom images are the reconstruction error maps of the corresponding methods.}\label{fig:paviax8}
\end{figure*}

\subsection{Results on Pavia Centre Dataset}
The Pavia Centre dataset is taken by Reflective Optics System Imaging Spectrometer (ROSIS) sensor, and it is a flight campaign over the center area of Pavia, northern Italy, in 2001. It has 102 spectral bands (the water vapor absorption and noisy spectral bands have been removed from the initially 115 spectral bands) and 1096$\times$1096 pixels in total. It should be noted that in the Pavia Centre scene, regions that contain no information are removed, leaving a meaningful region with 1096$\times$715 pixels.

To evaluate the proposed SSPSR method, we crop the center region of the image to obtain a subimage with 1096$\times$715 $\times$102 pixels, which is further divided into training and testing data. Specifically, the left part of this image are extracted to form the testing data, which has four non-overlap hyperspectral images with 223$\times$223 pixels. Besides, from the remaining region of the subimage, we extract overlap patches as reference high-resolution hyperspectral images for training (10\% of the training data is included as a validation set). Similar to previous settings, the patch size and low-resolution hyperspectral images are generated accordingly.

Table \ref{tab:Pavia} tabulates the average performance in terms of six PQIs over four testing images of all competing approaches. We can easily observe that the proposed SSPSR method significantly outperforms other algorithms with respect to almost all objective evaluation indexes. The average PSNR value of our method is 0.3 dB for $\times$4 and 0.2 dB for $\times$8 higher than the second best method. As the most competitive general gray/RGB image super-resolution methods, EDSR, RCAN, and SAN can achieve quite pleasurable results. However, their SAM indices are relatively poor when compared with these single hyperspectral image super-resolution methods, \emph{i.e.}, 3DCNN \cite{mei2017hyperspectral} and GDRRN \cite{li2018single}.

Fig. \ref{fig:paviax4} and Fig. \ref{fig:paviax8} show the reconstructed composite images and error maps of one test hyperspectral image in Pavia Center dataset of the six most competitive approaches with upsampling factors $d$ = 4 and $d$ = 8, respectively. The results of EDSR \cite{lim2017enhanced}, 3DCNN \cite{mei2017hyperspectral}, and GDRRN \cite{li2018single} are very blur, while RCAN \cite{zhang2018image} and SAN \cite{dai2019second} seem to introduce some noise. The proposed SSPSR method can maintain the main structural information. From the error maps of these methods, we can notice that the proposed method does not include obvious contour information of the image, which indicates that our method can well recover these information. It should be noted that when compared with the situation $d$ = 4, the visual results with upsampling factor $d$ = 8 are worse. In addition, when we compare the visual results of Fig. \ref{fig:japanx8} and Fig. \ref{fig:paviax8}, we also notice that reconstructed results on Pavia Center dataset are worse than these on Chikusei dataset. We think this is mainly due to the limited number of the training samples of the Pavia Center database. This is also a major drawback of these deep learning based methods. That is, they require a large number of training samples, otherwise they are difficult to train a model with promising generalization ability.

\begin{figure}[t]
  \centering
  \includegraphics[width=9cm]{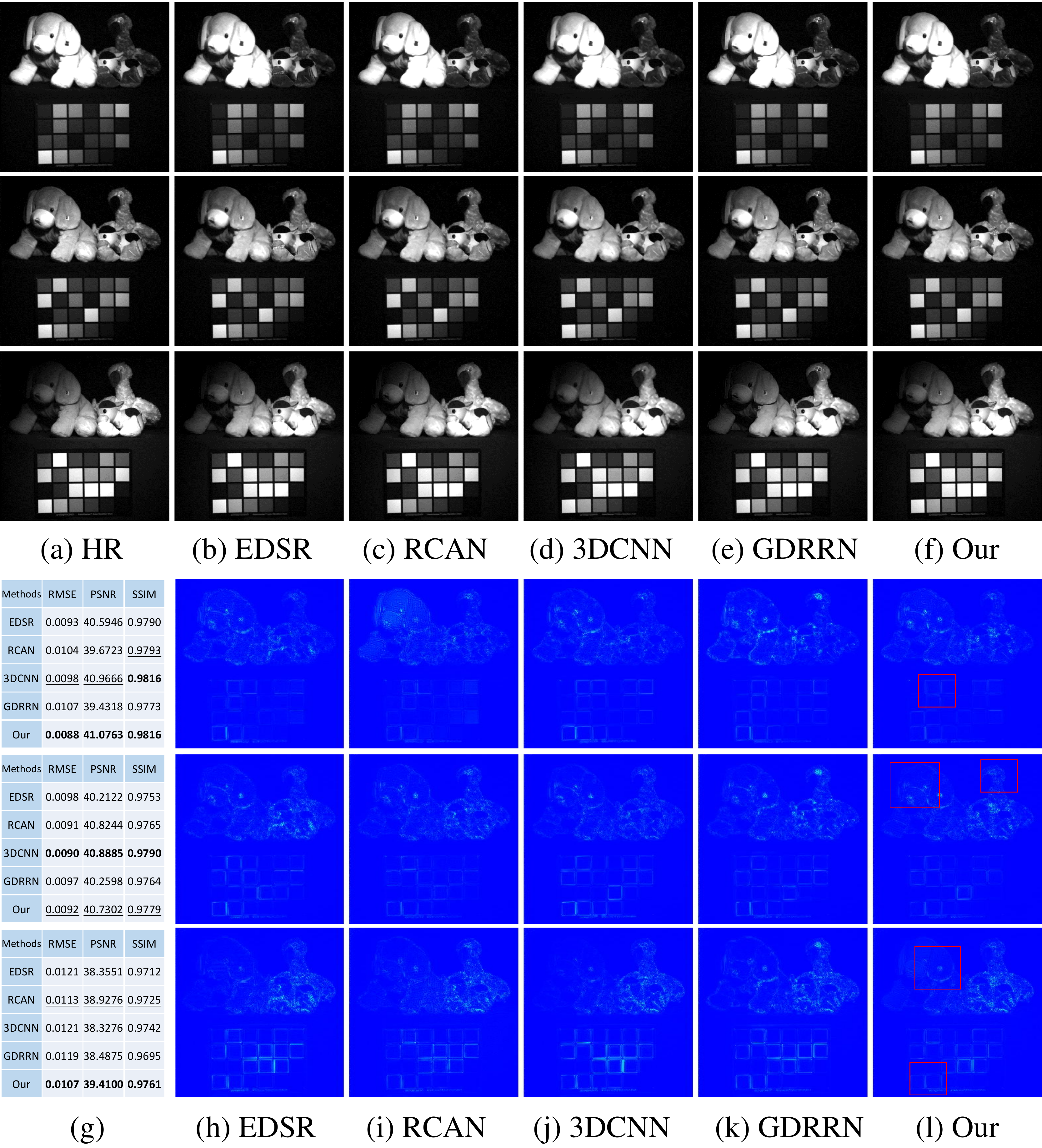}\\
  \vspace{-0.25cm}
  \caption{Reconstructed images of \emph{stuffed\_toys} at 480nm, 580nm and 680nm with upsampling factor $d$ = 4. The first 3 rows are the
reconstructed results for 480nm, 580nm and 680nm spectral bands, respectively; the last 3 rows show the error maps of the comparison methods. In (g), we report the RMSE, PSNR (dB), and SSIM results of the competing methods.}\label{fig:cavex4}
\end{figure}

\begin{figure}[t]
  \centering
  \includegraphics[width=9cm]{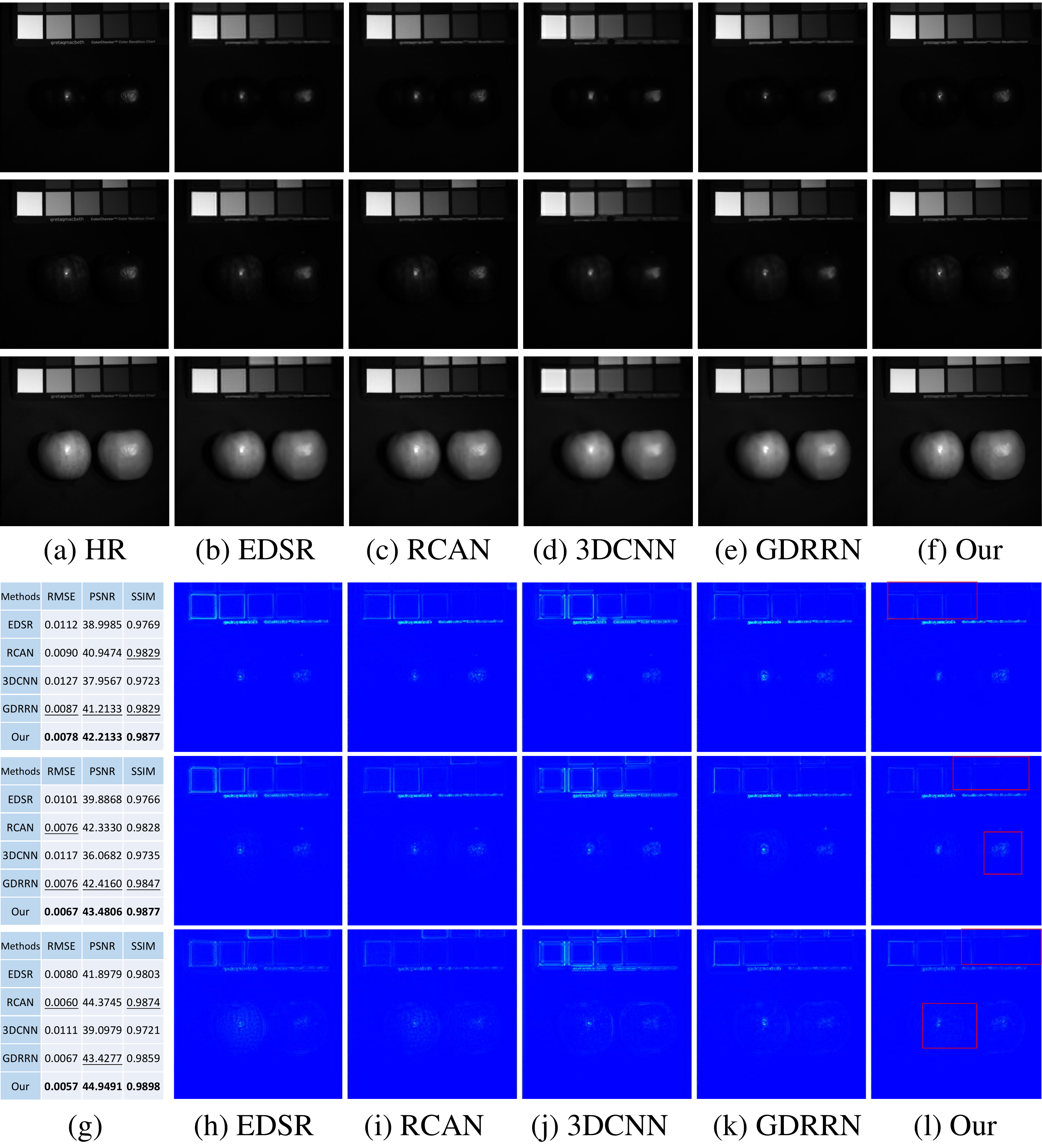}\\
    \vspace{-0.25cm}
  \caption{Reconstructed images of \emph{real\_and\_fake\_apples} at 480nm, 580nm and 680nm with upsampling factor $d$ = 8. The first 3 rows are the reconstructed results for 480nm, 580nm and 680nm spectral bands, respectively; the last 3 rows show the error maps of the comparison methods. In (g), we report the RMSE, PSNR (dB), and SSIM results of the competing methods.}\label{fig:cavex8}
\end{figure}

\subsection{Results on CAVE dataset}
The previous experiments are conducted on the Chikusei and Pavia Centre datasets, which are all remotely sensed hyperspectral images. To further verify the effectiveness of the proposed SSPSR method, we also conduct comparison experiments on hyperspectral images of natural scenes. Specifically, we use the CAVE multispectral image database because it is widely used in many multispectral image recovery tasks.
The database consists of 32 scenes of everyday objects with spatial size of 512$\times$512, including 31 spectral bands ranging from 400nm to 700nm at 10nm steps. To prepare samples for training, we randomly select 20 hyperspectral images from the database (10\% samples are randomly selected for evaluations). When the upsampling factor $d$ is 4, we extract patches with 64$\times$64 pixels (32 pixels overlap) for training; when the upsampling factor $d$ is 8, we let the extracted patches as 128$\times$128 pixels (with 64 pixels overlap). The corresponding low-resolution hyperspectral image are generated by Bicubic downsampling with a factor of 4 or 8. The remaining 12 hyperspectral images of the database are used for testing, where the original images are treated as ground truth high-resolution hyperspectral images, and the low-resolution hyperspectral inputs are generated similarly as the training samples. For this dataset, we set the spectral band number ($p$) of each group to 4 and the overlap ($o$) between neighboring groups to 1. Since the Cave dataset can provide more training samples, we use a larger $R (R=8)$ to design our network.

We compare the proposed SSPSR method with some very competitive approaches, EDSR \cite{lim2017enhanced}, RCAN \cite{zhang2018image}, 3DCNN \cite{mei2017hyperspectral}, and GDRRN \cite{li2018single}. The average performance of the CC, SAM, RMSE, ERGAS, PSNR, and SSIM results of competing methods for different upsampling factors on the CAVE dataset are reported in Table \ref{tab:Cave}. From these results, we notice that the 3DCNN method performs worse than other methods. Clearly, the proposed SSPSR method outperforms all other competing methods. The proposed SSPSR method performs much better than EDSR \cite{lim2017enhanced} and RCAN \cite{zhang2018image}, which focus on exploiting the spatial prior. On average, the PSNR and SSIM values of the proposed SSPSR method for upsampling factor $d$ = 4/8 are 0.3/0.4 dB and 0.002/0.012 higher than the second best method, respectively.

Fig. \ref{fig:cavex4} and Fig. \ref{fig:cavex8} show the reconstructed HR hyperspectral images and the corresponding error maps at 480nm, 580nm and 680nm by the competing methods for test images \emph{stuffed\_toys} and \emph{real\_and\_fake\_apples} with upsampling factors $d$ = 4 and $d$ = 8, respectively. From the visual reconstruction results, we can see that all the comparison methods can well reconstruct the high-resolution spatial structures of the hyperspectral images. In these error maps, we learn that the proposed method and RCAN method achieve the best reconstruction fidelity in recovering the details of the original hyperspectral images. For example, the edges of the checkerboards and the contours of dog's ears and apples (please refer to the regions marked with red boxes). In the subfigure (g), we also report the RMSE, PSNR, and SSIM results of each spectral band for the competing methods. Obviously, the proposed SSPSR method performs best in most cases. 3DCNN \cite{mei2017hyperspectral} and GDRRN \cite{li2018single}, which are designed for the hyperspectral images, can achieve favorable results in some cases, but their performance seems to be unstable when reconstructing different spectral bands.

\begin{table}
%\vspace{1mm}
\begin{center}
\caption{Quantitative comparisons of different approaches over 12 testing images from Cave dataset with respect to six PQIs. }\label{tab:Cave}
%\vspace{1.5mm}
%\footnotesize
%\setlength{\tabcolsep}{7.5pt}
\begin{tabular}{cccccccc}

  \hline
	&	$d$	&	CC$\uparrow$	&	SAM$\downarrow$	&	RMSE$\downarrow$	&	ERGAS$\downarrow$	&	PSNR$\uparrow$	&	SSIM$\uparrow$	\\
    \hline
        \hline
Bicubic	&	4	&	0.9868 	&	4.1759 	&	0.0212 	&	5.2719 	&	34.7214 	&	0.9277 	\\
EDSR \cite{lim2017enhanced}	&	4	&	0.9931 	&	3.5499 	&	0.0149 	&	3.5921 	&	38.1575 	&	0.9522 	\\
RCAN \cite{zhang2018image}	&	4	&	\underline{0.9935}	&	3.6050 	&	\underline{0.0142}	&	\underline{3.4178}	&	\underline{38.7585}	&	0.9530 	\\
SAN \cite{dai2019second}	&	4	&	\underline{0.9935}	&	3.5951 	&	0.0143 	&	3.4200 	&	38.7188 	&	0.9531 	\\
3DCNN \cite{mei2017hyperspectral}	&	4	&	0.9928 	&	\underline{3.3463}	&	0.0154 	&	3.7042 	&	37.9759 	&	0.9522 	\\
GDRRN \cite{li2018single}	&	4	&	0.9934 	&	3.4143 	&	0.0145 	&	3.5086 	&	38.4507 	&	\underline{0.9538}	\\
SSPSR	&	4	&	\textbf{0.9939}	&	\textbf{3.1846}	&	\textbf{0.0138}	&	\textbf{3.3384}	&	\textbf{39.0892}	&	\textbf{0.9553}	\\
    \hline
    \hline
Bicubic	&	8	&	0.9666 	&	5.8962 	&	0.0346 	&	4.2175 	&	30.2056 	&	0.8526 	\\
EDSR \cite{lim2017enhanced}	&	8	&	0.9778 	&	5.6865 	&	0.0279 	&	3.3903 	&	32.4072 	&	0.8842 	\\
RCAN \cite{zhang2018image}	&	8	&	0.9791 	&	5.9771 	&	0.0268 	&	3.1781 	&	32.9544 	&	0.8884 	\\
SAN \cite{dai2019second}	&	8	&	\underline{0.9795}	&	5.8683 	&	\underline{0.0267}	&	\underline{3.1437}	&	\underline{33.0012}	&	0.8888 	\\
3DCNN \cite{mei2017hyperspectral}	&	8	&	0.9755 	&	\underline{5.0948}	&	0.0292 	&	3.5536 	&	31.9691 	&	0.8863 	\\
GDRRN \cite{li2018single}	&	8	&	0.9769 	&	5.3597 	&	0.0280 	&	3.3460 	&	32.5763 	&	\underline{0.8890}	\\
SSPSR	&	8	&	\textbf{0.9805}	&	\textbf{4.4874}	&	\textbf{0.0257}	&	\textbf{3.0419}	&	\textbf{33.4340}	&	\textbf{0.9010}	\\
  \hline
\end{tabular}
\normalsize
\end{center}
\end{table}

\section{Conclusions}
\label{sec:Conclusions}
In this paper, a novel deep neural network based on spatial-spectral prior network (SSPN) is introduced to address the single hyperspectral image super-resolution problem. In particular, in order to discover the spatial and spatial correlation characteristics of hyperspectral data, we carefully designed a spatial-spectral prior network (SSPN) to fully exploit the spatial information and correlation among the different spectral features. In addition, to cope with the problems that the training samples of hyperspectral image are limited and the dimensionality is high, a group convolution (with shared network parameters) and progressive upsampling framework is proposed. In this way, we can expect to greatly reduce the parameters of the model and make it possible to obtain stable training results under small data and large spectral band number conditions. In our introduced network, the transmission of information flow is very flexible by the short, long, global skip links via residual learning. To regularize the network outputs, we adopt a spatial-spectral total variation (SSTV) based constraint to preserve the edge sharpness spectral correlations of the super-resolved high-resolution hyperspectral image. Evaluations on three public hyperspectral datasets demonstrate that our model not only achieves the best performance in terms of some commonly used objective indicators, but also generates clear high-resolution images which are perceptually closer to the ground truth when compared with state-of-the-arts.

%\section{Acknowledgment}
%The authors would like to thank Dr. Yicong Zhou from University of Macau for sharing his algorithm codes for comparison purposes.

\ifCLASSOPTIONcaptionsoff
  \newpage
\fi



%\bibliographystyle{IEEEtran}
{
\bibliographystyle{IEEEtran}
\bibliography{RLPA2018}
}

%
%\begin{IEEEbiography}[{\includegraphics[width=1.0in,height=1.25in,clip,keepaspectratio]{Chen}}]{Chen Chen}
%received the B.E. degree from Beijing Forestry University, Beijing, China, in 2009 and the MS degree from Mississippi State University, Starkville, in 2012 and the PhD degree from the University of Texas at Dallas, Richardson, TX in 2016.
%
%From 2016 to 2018, he was a Post-Doc in the Center for Research in Computer Vision at University of Central Florida (UCF). He is currently an Assistant Professor with Department of Electrical and Computer Engineering, University of North Carolina at Charlotte. His research interests include compressed sensing, signal and image processing, pattern recognition and computer vision. He has published over 40 papers in refereed journals and conferences in these areas.
%\end{IEEEbiography}

\end{document}